\documentclass[conference,a4paper]{IEEEtran}
\usepackage{cite}
\usepackage{amsmath,amssymb,amsfonts}
\usepackage{algorithmic}
\usepackage{graphicx}
\usepackage{textcomp}
\usepackage{subfig}
\usepackage{float} 
\usepackage{xcolor}
\usepackage{caption} 
\usepackage{subcaption} 
\usepackage{mathtools}

\def\BibTeX{{\rm B\kern-.05em{\sc i\kern-.025em b}\kern-.08em
    T\kern-.1667em\lower.7ex\hbox{E}\kern-.125emX}}
\begin{document}

\title{Experimental Evaluation of Data Upload Efficiency and Guiding Challenges for a Vehicular-to-Road System Using 60-GHz mmWave Ultra-Spots
\thanks{}
}

\author{\IEEEauthorblockN{Phuc Duc Nguyen\IEEEauthorrefmark{1}, Kazuhiro Maruyama\IEEEauthorrefmark{1}, Keitarou Kondou\IEEEauthorrefmark{1}\IEEEauthorrefmark{2}, Yozo Shoji\IEEEauthorrefmark{1}}
\IEEEauthorblockA{\textit{\IEEEauthorrefmark{1}Social-ICT System Laboratory, National Institute of Information and
Communications Technology (NICT), Tokyo, Japan}\\
\textit{\IEEEauthorrefmark{2}HRCP Research and Development Partnership, Tokyo, Japan}\\
Corresponding emails: phucdnguyen@nict.go.jp (P. D. N.), and shoji@nict.go.jp (Y. S.)}
}

\maketitle

\begin{abstract}
Maximizing data uploading efficiency in a vehicular-to-road data uploading system using millimeter-wave communication is a challenging issue, as the wireless zone is often critically narrow, and vehicles can easily fail to pass through it without the aid of an autonomous guiding system. Variations in driving routes, speeds, approach angles, and distances to the ultra-spot can significantly affect data transmission performance, leading to either efficient or suboptimal results. This study presents a comprehensive analysis based on 75 experimental cases to identify the optimal travel trajectory and conditions that allow the vehicle to pass through the ultra-spot and enhance data transmission effectively. Experimental results show that with an optimal travel trajectory, appropriate movement speed, antenna placement, and prior estimation of the ultra-spot area, the amount of transferred data can be improved by 6 to 8 times.
\end{abstract}


\section{Introduction and Methodology}
\label{sec:01}

In pursuing ultra-high-speed and ultra-reliable mobile communication in Beyond 5G (B5G) systems, the concept of ultra-spots has emerged as a novel solution\cite{nguyen2025prediction}. Ultra-spots are spatially limited wireless zones that provide ultra-high-capacity, low-latency connectivity to pre-authorized users. They are designed for highly localized and secure communication, enabling personalized data services tailored to individual needs, available \textit{only now}, \textit{only here}, and \textit{only for specific users}. However, since their wireless coverage is extremely narrow, various challenges arise when they are utilized by mobile entities such as vehicles or UAVs. Such zones may be deployed at selected locations where conventional infrastructure, such as local 5G base stations, is either too costly or impractical to maintain, e.g., roadside stations, government facilities, or mobile event hubs\cite{yue2019millimeter}. These ultra-spots are especially promising for burst-mode, high-volume data uploading in vehicular or UAV scenarios\cite{liu2025measurements}. However, their narrow spatial footprint imposes significant challenges regarding real-time mobility trajectory prediction, antenna alignment, and communication continuity. Several technical challenges arise in this context, including accurately predicting and guiding mobile platforms toward ultra-spots, establishing reliable short-duration links, and evaluating data throughput and uploading efficiency in such dynamic environments. This paper addresses these challenges by analyzing real-world vehicle behavior in passing through ultra-spot zones, based on field experiments conducted in March 2025 at Michi-no-Eki Kokindenju-no-Sato Yamato, Gifu, Japan. A prototype system with a fixed roadside receiver and a mobile mmWave transmitter on the vehicle was used to evaluate high-capacity data uploading. As shown in Fig.\,\ref{fig:01}, 75 different experimental cases were conducted, varying in drivers, routes, speeds, antenna placements, and whether or not a buzzer was used to notify the approach to the ultra-spot. Both ends were equipped with mmWave transceivers transmitting at –5\,dBm and horn antennas with a gain of 24\,dBi. The data uploading performance was recorded in each experiment, helping to identify the optimal trajectory and conditions to maximize communication throughput at the ultra-spot.

\section{Results and Conclusions}

 \begin{figure}[!t]
	\centerline{\includegraphics[width=1\columnwidth]{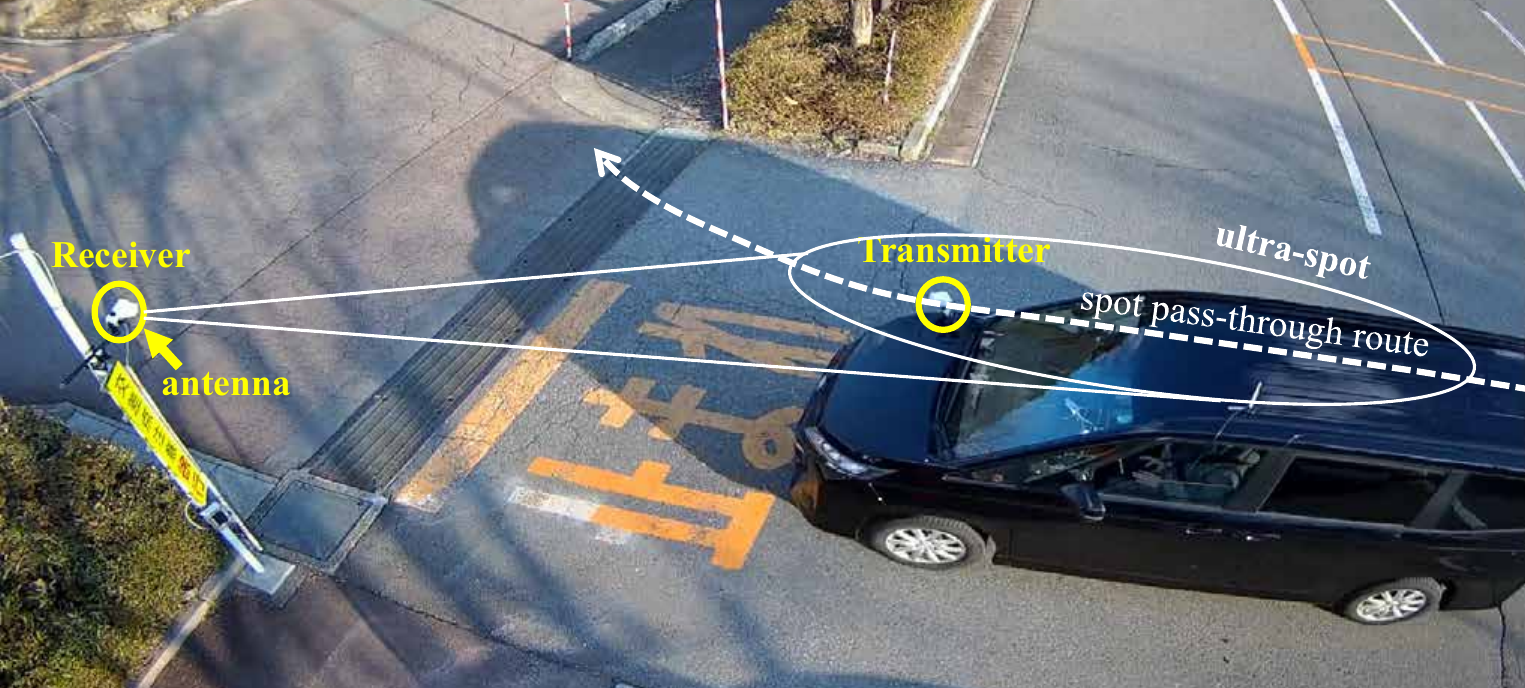}}
	\caption{An experimental setup and a passing-through route illustration for the vehicular-to-road data uploading system.}
	\label{fig:01}
\end{figure}

Fig.\,\ref{fig:rssi} shows a 2D RSSI heatmap of the mmWave ultra-spot coverage area, generated by overlaying RSSI data from the mmWave transceiver across 20 experimental trials with driver A using an external antenna and buzzer-assisted approach. The experimental area, which estimates the coverage of the ultra-spot, as shown in Fig.\,\ref{fig:01}, is divided into multiple rectangular cells on the X-Y plane, as shown in Fig.\,\ref{fig:rssi}. Each cell in the heatmap has dimensions of 25.4 cm x 40.1 cm. Let the parameters be defined as follows: $G_r$ and $G_t$ represent the maximum gains of the receiving and transmitting antennas, respectively. The directional gain patterns of these antennas are denoted by $A_r(\theta, \alpha, \gamma)$ for the receiver and $A_t(\theta, \alpha, \gamma)$ for the transmitter, where $\theta$, $\alpha$, and $\gamma$ indicate angular parameters. The distance between the transmitter (TX) and receiver (RX) is denoted as $d$, and the free-space path loss over this distance is represented by $L(d)$. The received power $P_r$ at a point $(x, y)$ on the $X,Y$ plane (interpreted as RSSI in Fig.\,\ref{fig:rssi}) can be approximated as:

\begin{figure}[!t]
	\centerline{\includegraphics[width=1\columnwidth]{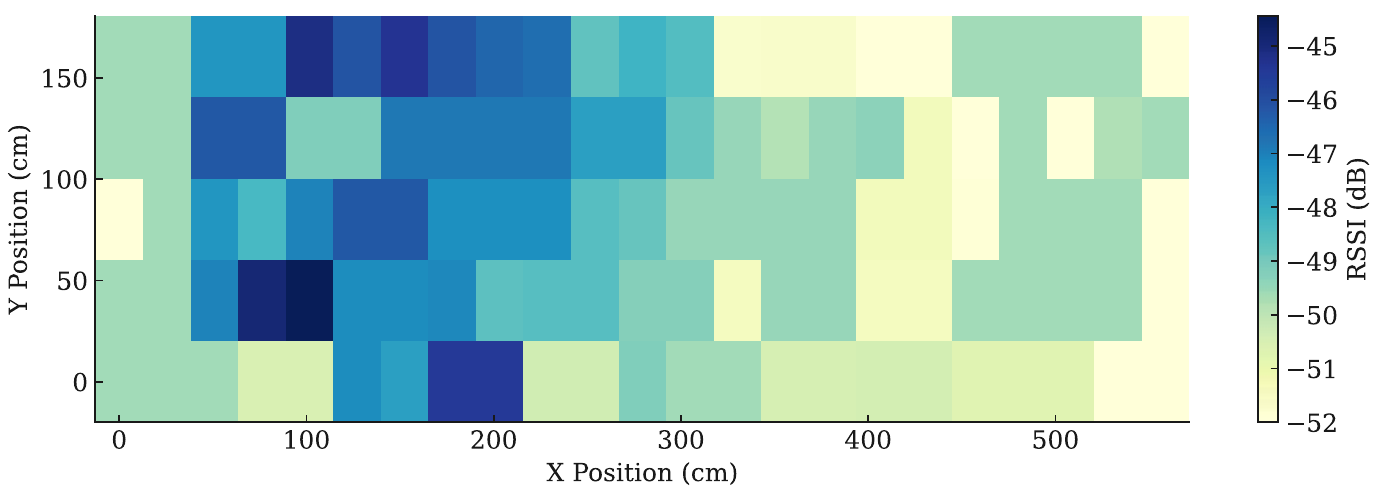}}
	\caption{Estimated ultra-spot zone visualized as 2D RSSI heatmap.}
	\label{fig:rssi}
\end{figure}

\begin{equation}
P_r(x, y) \propto \frac{G_t \cdot A_t(\theta, \alpha, \gamma) \cdot G_r \cdot A_r(\theta, \alpha, \gamma)}{L(d)}
\end{equation}

where the path loss function is defined by: $L(d) = \left( \frac{4\pi d}{\lambda} \right)^2$; here, $\lambda$ is the wavelength of the signal. 

 \begin{figure}[!t]
	\centerline{\includegraphics[width=1\columnwidth]{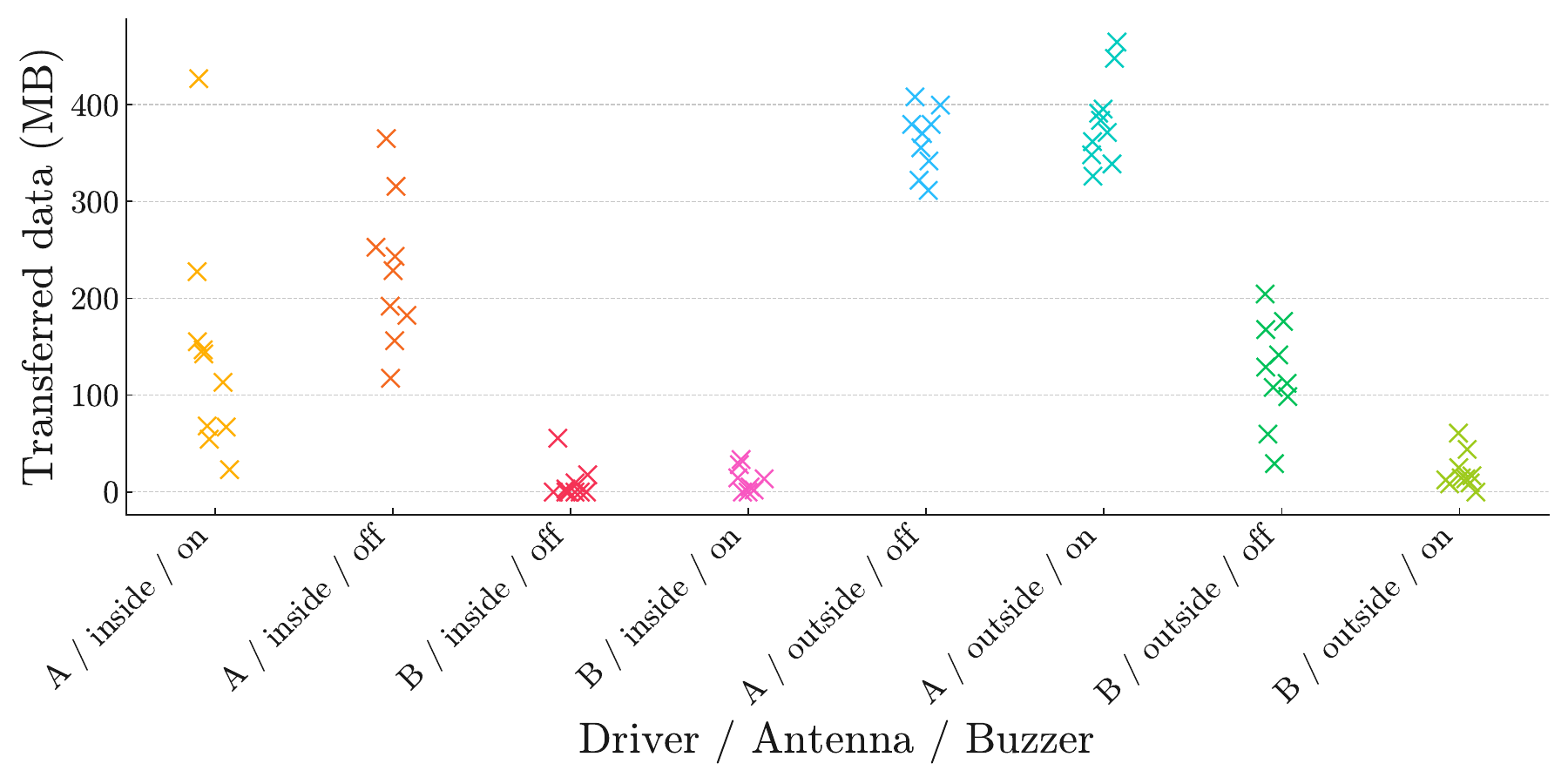}}
	\caption{The volume of data transmitted from a moving vehicle to a roadside station was evaluated under eight distinct experimental conditions.}
	\label{fig:02}
\end{figure}

 \begin{figure}[!t]
	\centerline{\includegraphics[width=1\columnwidth]{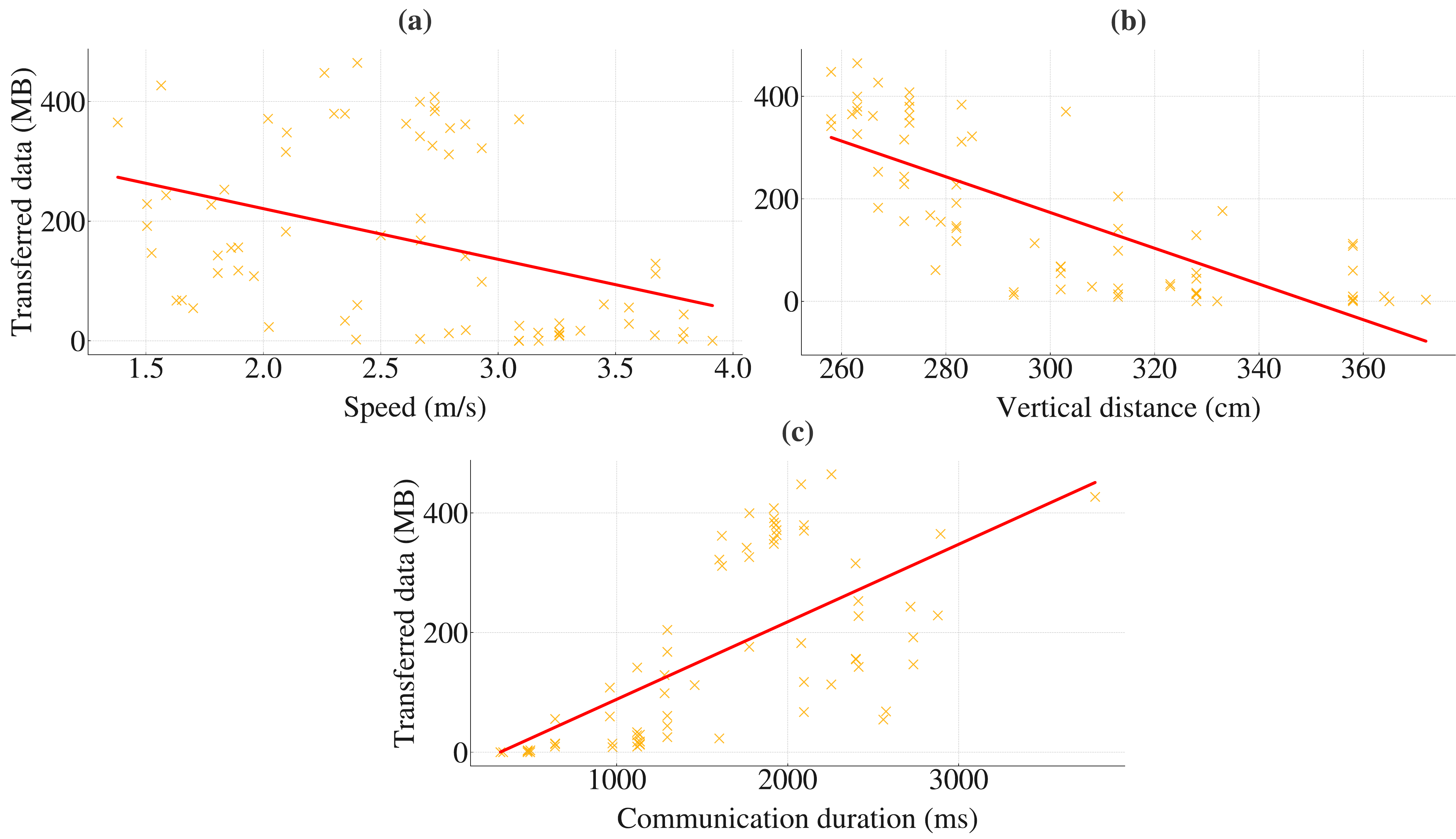}}
	\caption{The data volume transmitted within the ultra-spot was analyzed in relation to (a) vehicle speed, (b) TX/RX antenna separation, and (c) transmission duration.}
	\label{fig:03}
\end{figure}

As shown in Fig.\,\ref{fig:02}, data transmission from a moving vehicle to a roadside station was evaluated under eight conditions, varying by driver (A or B), antenna placement (inside or outside the vehicle), and presence or absence of a buzzer indicating link establishment. It can be seen that most of the highest data transmission cases (over 300\,MB) were achieved by driver A, using a buzzer and an mmWave antenna mounted outside the vehicle. Conversely, for driver B, who had a different speed and route compared to driver A, and with the antenna mounted inside the car and no buzzer used, the amount of data transmitted to the roadside station decreased. These findings highlight the importance of driving behavior, speed, trajectory, and timely ultra-spot detection in efficient data transmission.

Figure\,\ref{fig:03}(a) illustrates the relationship between the vehicle’s speed when passing through the ultra-spot and the amount of data transmitted to the roadside station across 75 repeated test runs. It can be observed that the general trend is that as the speed increases, the amount of transmitted data decreases. The ideal speed range for passing through the ultra-spot is between 1.9 m/s and 2.5 m/s. Fig.\,\ref{fig:03}(b) illustrates the relationship between the vertical distance (top-down view) between the TX antenna (mounted on the vehicle) and the RX antenna (mounted at the roadside station) throughout the vehicle's movement. Across 75 repeated experiments, the general trend shows that shorter vertical distances result in more data being transmitted, with the optimal range being between 260\,cm and 280\,cm. The reason we did not test distances shorter than 260\,cm is due to safety considerations: a required safety margin of approximately 1.3\,m between the vehicle and the roadside station, along with a 1.3\,m distance from the outermost left edge of the vehicle to the antenna position, totaling a minimum of 2.6\,m.  Fig.\,\ref{fig:03}(c) illustrates the relationship between the actual data transmission time as the vehicle passes through the ultra-spot and the amount of data transmitted to the roadside station across 75 repeated test runs. It can be seen that a longer transmission time does not necessarily mean more data is transmitted in every case, although that is the general trend. The ideal transmission time is between 1.8 and 2.3 seconds, which over 400\,MB of data was successfully transmitted within this short duration.

Based on experimental results, the optimal conditions for maximizing ultra-spot data transfer efficiency are: a vertical distance of 260–280\,cm, speed of 1.9–2.5\,m/s, and transmission time of 1.8–2.3\,s. The antenna should be mounted outside the vehicle, and ultra-spot ideally detected in advance. These insights support future development of autonomous route adjustment systems to maximize roadside data uploading.

\bibliographystyle{IEEEtran}

\begin{thebibliography}{1}
\providecommand{\url}[1]{#1}
\csname url@samestyle\endcsname
\providecommand{\newblock}{\relax}
\providecommand{\bibinfo}[2]{#2}
\providecommand{\BIBentrySTDinterwordspacing}{\spaceskip=0pt\relax}
\providecommand{\BIBentryALTinterwordstretchfactor}{4}
\providecommand{\BIBentryALTinterwordspacing}{\spaceskip=\fontdimen2\font plus
\BIBentryALTinterwordstretchfactor\fontdimen3\font minus \fontdimen4\font\relax}
\providecommand{\BIBforeignlanguage}[2]{{%
\expandafter\ifx\csname l@#1\endcsname\relax
\typeout{** WARNING: IEEEtran.bst: No hyphenation pattern has been}%
\typeout{** loaded for the language `#1'. Using the pattern for}%
\typeout{** the default language instead.}%
\else
\language=\csname l@#1\endcsname
\fi
#2}}
\providecommand{\BIBdecl}{\relax}
\BIBdecl

\bibitem{nguyen2025prediction}
P.~Nguyen \emph{et~al.},``Prediction of Ultra-high-speed Spots Using RTK-GNSS Sensor Fusion for UAV-to-UAV mmWave/THz Communications'', in \emph{IEEE Access} (2025).

\bibitem{yue2019millimeter}
Y.~Guangrong \emph{et~al.},``Millimeter-wave system for high-speed train communications between train and trackside: System design and channel measurements'', in \emph{IEEE Transactions on Vehicular Technology}.\hskip 1em plus 0.5em minus 0.4em\relax vol. 68, no. 12, pp.11746-11761, 2019. 

\bibitem{liu2025measurements}
L.~Xichen \emph{et~al.},``Measurements and Modeling of Millimeter-Wave Vehicle-to-Vehicle Propagation With Vehicle Obstructions'', in \emph{IEEE Transactions on Wireless Communications}\hskip 1em plus 0.5em minus 0.4em\relax (2025).

\end{thebibliography}


\end{document}